\documentclass[aip,jmp,amsmath,amssymb,reprint]{revtex4-1}
\usepackage{graphicx}
\usepackage{dcolumn}
\usepackage{bm}
\begin{document}

\preprint{AIP/123-QED}

\title[Sample title]{ Studying the applicability of different thermoelectric materials for efficiency calculation in hybrid thermoelectric generator for waste heat recovery from automobile and steel industry}
\author{Kumar Gaurav}
\email{kumargauravmenit@gmail.com}
 \altaffiliation[also at]{ School of Engineering. Indian Institute Of Technology Mandi, Kamand, Himachal Pradesh, India, 175005}
\author{Sudhir K Pandey}
 \email{sudhir@iitmandi.ac.in}
\affiliation{ School Of Engineering, Indian Institute Of Technology Mandi, Kamand, Himachal Pradesh, India, 175005} 
\date{\today}

\begin{abstract}
In this work, we study the suitability of different thermoelectric materials like $Bi_2Te_3$, $Sb_{2}Te_{3}$, $PbTe$, $TAGS$, $CeFe_{4}Sb_{12}$, $SiGe$ and $TiO_{1.1}$ in estimation of thermoelectric generator's (TEG) efficiency. The efficiency of TEG made up of ${Be_{2}Te_{3}}$ or $Sb_{2}Te_{3}$ gives $\sim$7\% in temperature range of 310 K - 500 K. $PbTe$ or $TAGS$ or $CeFe_{4}Sb_{12}$ gives $\sim$6\% in temperature range of 500 K - 900 K and $SiGe$ or $TiO_{1.1}$ also have remarkable efficiency in higher temperature range i.e $\sim$1200 K. Here, we report the enhancement of efficiency by using hybridization technique for different combination of above-mentioned materials. Hybridization of two different materials of TEG module is done by considering compatibility factor aspect. To this end, the proposed values of overall efficiency of TEG by hybridizing ${Be_{2}Te_{3}}$ and $PbTe$; ${Be_{2}Te_{3}}$ and $TAGS$; $Bi_{2}Te_{3}$ and $CeFe_{4}Sb_{12}$ are 12\%, 14\% and 11.88\%, respectively, for temperature range of 310 K to 900 K, which can be installed in an automobile. For steel industry and spacecraft application (till 1200 K) hybridization of $Bi_{2}Te_{3}$, $PbTe$ and $SiGe$; ${Be_{2}Te_{3}}$ and $TiO_{1.1}$ yields efficiency of $\sim$15.2\% and $\sim$17.2\%, respectively. The proposed results can be treated as a viable option for engineers, who are looking for fabricating TEG in real life applications such as automobile, spacecraft and steel industry.
\end{abstract}
\maketitle
\bigskip

\thispagestyle{plain}
\pagestyle{plain}
\section*{INTRODUCTION}
Extraction of energy through TEG plays an important role in reducing our dependencies over the non-renewable source of energy. Installation of TEG can be done in numerous places such as general and special types of automobile's exhaust\cite{Fair} (usually, stryker combat vehicle). Spacecraft organization can use as General Purpose Heat Source\cite{Spacecraft} (GPHS) radioisotopes TEG (fueled by Pu-238). Similarly, for small scale applications, TEG can be installed in a wristwatch\cite{G Jeffrey}, man-portable power, etc. Large scale production can be achieved by installing TEG in power plant exhaust\cite{India} (the main source of power production in the world) as well as steelworks industry\cite{Firoz}, where an enormous amount of energy is being wasted out. The energy obtained through TEG setup can be utilized to satisfy our auxiliary needs\cite{Spacecraft}, such as belt driven accessories can be replaced by electric motor, charging the battery, electrical power requirement of vehicle, telematics (GPS), collision avoidance system, navigation systems, electronic braking, powertrain body controllers and sensors, LED, etc. By applying the same concept of TEG, other application of thermoelements can be as refrigeration purpose.\par In spite of great advantages and scope of TEG, we have certain challenges regarding theoretical calculation of efficiency and installation of TEG. Firstly, we should calculate theoretically the accurate efficiency of TEG, secondly, to enhance the efficiency within a specific temperature range. Other complementary necessities for good TEG operations are 1. It should be strong enough to sustain in the vibrational environment (automobile), 2. The materials of TEG should have good heat and oxidation resistance, 3. High energy conversion efficiency, and 4. Low production cost.\par As the first challenge is concerned with efficiency calculation, which is already taken care in our previous work\cite{Gaurav}. Since, many research groups experimentally found the value of thermal conductivity ($K$), electrical resistivity ($\rho$) and Seebeck coefficient($S$) of numerous thermoelectric materials. So, in order to use that results, we have already formulated a method to calculate the efficiency of TEG (which can be installed in any of the above-mentioned application sites) in our previous work\cite{Gaurav}. Despite great research and development in the field of TEG, the efficiency is nearly 7\%, which is quite low. The parallel connection of thermoelectric materials in TEG module leads to increase the power output, however, the operating temperature range gets restricted. So, the second challenge is, to enhance the efficiency of TEG. This increment in efficiency can be achieved by using a thermoelectric material having a good figure of merit ($z\bar{T}$) within entire working temperature range and which can sustain at higher temperature. Till now, it is very difficult to attain these two favorable conditions because a single material does not have higher $z\bar{T}$ in a broad temperature range as well as its suitability in higher temperature. So, for solving these complexities, we are using hybrid technology, in which different thermoelectric materials are stacked in series connection and net output is being taken between two extreme ends of the entire module. Selection of different materials are done on the basis of their good figure of merit in different temperature range and the highest temperature is decided by their melting point and other physical properties. Accordingly different materials are chosen for different temperature range and stacking one over other is done. Now, one more problem got emerged in the form of working temperature range for all individual segment of the hybrid module, means problem regarding interface temperature. The aim is, to choose a specific temperature of the cold end of the top layer and hot end of the bottom layer to maximize efficiency. That interface temperature is decided by compatibility factor\cite{Snyder2003}, where compatibility factor is defined as the ratio of current density and heat flux (compatibility factor at the interface of both layers should be close enough to maximize efficiency). Other challenges like high mechanical strength, good heat and oxidation resistance (nowadays oxide becomes the focus of attention for high-temperature application), high energy conversion efficiency, low production cost, etc., are taken care by individual examination of each material through reports presented by the different research groups.\par Here we report the efficiency of TEG using different individual thermoelectric materials within their favorable temperature range. We found out the calculated efficiency of ${Be_{2}Te_{3}}$ and $Sb_{2}Te_{3}$ in the lower temperature range (310 K to 570 K) is around 7.20\%. Similarly, efficiency of TEG using $PbTe$, $TAGS$ ($(AgSbTe_{2})_{0.15}(GeTe)_{0.85}$), $CeFe_{4}Sb_{12}$ individually in medium temperature range (500 K to 900 K) is found out around 6\% to 8\%. For TEG made up of$SiGe$ or $TiO_{1.1}$ in higher temperature range (up to $\sim$1200 K) is also having comparable efficiency. We also found out efficiency by stacking (hybrid) different combinations of above materials, which gives around 12\% for ${Be_{2}Te_{3}}$ and $PbTe$ as well as around 14\% for ${Be_{2}Te_{3}}$ and $TAGS$, etc. Consequently, the applicability of hybrid TEG is discussed in automobiles as well as steel industry. Considering fixed hot end temperature, we came up with temperature dependent efficiency for a combination of different compounds as a hybridized module. Efficiency obtained for different application sites lies in the range of 12-17\%, which is significantly good output. 
\vspace{-0.5cm}
\section*{METHODOLOGY}
Our previous work\cite{Gaurav} deals with efficiency calculation of TEG made up of a single thermoelement, where, segmentation of entire sample is done in order to calculate the efficiency of TEG operating in a particular temperature range. This segmentation process will make $z\bar{T}$ as a temperature independent identity. Considering, Seebeck effect, Joule's effect, and Thomson effect, involve in a thermoelectric material, we have the formula for calculating maximum efficiency of a TEG\cite{Rowe}, \cite{Roy Taylor} as:
\begin{equation}
\centering
\eta_{max}={\frac{T_{h}-T_{c}}{T_{h}}}\frac{\sqrt{1+z\bar{T}}-1}{\sqrt{1+z\bar{T}}+\frac{T_{c}}{T_{h}}}
\end{equation} 
where, $\bar{T}=\frac{T_{h}+T_{c}}{2}$, $T_{h}$, $T_{c}$ and $z\bar{T}$ denote average, source, sink temperature and figure of merit, respectively. Correspondingly, we have calculated the efficiency of TEG for all individual segments. Thereafter, by using simple thermodynamics formulation\cite{L Chen}, \cite{Lihua Zhang} as shown in Eq.(2), we found out the overall efficiency of TEG made up of particular thermoelectric material.
\begin{equation}
\centering
\eta_{overall} = 1-(1-\eta_{1})(1-\eta_{2})(1-\eta_{3})......(1-\eta_{n})
\end{equation} where, $\eta_{1}$, $\eta_{2}$, $\eta_{3}$....$\eta_{n}$ are the individual efficiency of segmented sections $1^{st}$, $2^{nd}$, $3^{rd}$, $n^{th}$, respectively. On the continuation of our previous work\cite{Gaurav}, here we applied the series combination of different thermoelectric material to increase efficiency. Calculation of thermoelectric efficiency followed the process of segmentation of one thermoelement into a number of small pieces such as segment 1,2,3...so-on and similarly for another thermoelement as shown in the FIG.1.
\begin{figure}[ht!]
\centering
\includegraphics[width=1.0in]{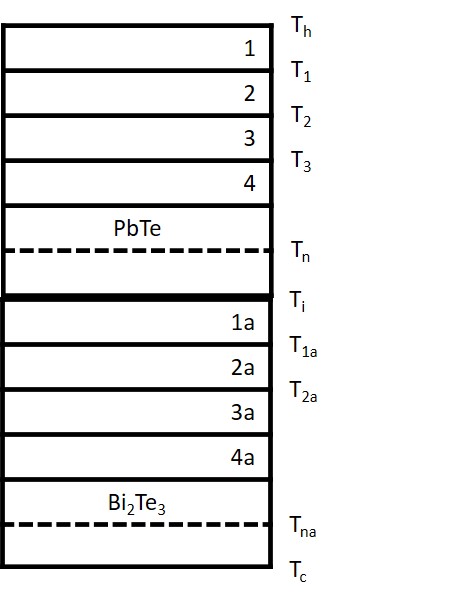}
\caption{Schematic of segmentation of hybridized TEG comprising of thermoelement 1 and 2}
\end{figure} We have calculated the efficiency of both top and bottom stacked layers for temperature range of $T_{h}$ to $T_{i}$ and $T_{i}$ to$ T_{c}$, respectively. It gives efficiency for two layers as $\eta_{overall1}$ and $\eta_{overall2}$. On considering the ideal situation, where, there is no heat loss, we can calculate the overall efficiency of TEG module by using the same formulation as explained for the segmented system\cite{LJ Ybarrondo}:
\begin{equation}
\centering
\eta_{overall}=1-(1-\eta_{overall1})(1-\eta_{overall2})
\end{equation}. This formula gives the cumulative efficiency for different combinations of thermoelectric materials i.e $\eta_{overall}$ for a temperature range of $T_{h}$ to $T_{c}$. This procedure is quite simple for calculating TEG efficiency because only by knowing $z\bar{T}$ (a material property which depends on $S$, $k$ and $\rho$), $T_{h}$ and $T_{c}$ one can find the efficiency of specific TEG module. Our next problem is about deciding the working temperature range for individual thermoelectric material. For stacking two different materials in series connection, we have to search the best suitable temperature at the interface surface for maximum efficiency. That is done by considering compatibility factor $u$, which is defined as the ratio of current density to heat flux. It tells that for $u =\frac{\sqrt{1+z\bar{T}}-1}{S\bar{T}}$ will gives us maximum efficiency, where $u$ and $S$ are compatibility factor (relative current density) and Seebeck coefficient, respectively. Here, the value of $u$ is obtained by maximizing the reduced efficiency\cite{Ursell} formula with respect to $u$, keeping other parameters as constant.
\begin{equation}
\eta_{r}=\frac{u\frac{S}{z}(1-u\frac{S}{z})}{u\frac{S}{z}+\frac{1}{z\bar{T}}}
\end{equation} Generally hybridization of two layers are fruitful only when, compatibility factor of both layers at interface temperature should not differ by a factor of 2 or more. So, our aim is to search different materials which should be compatible with each other and their applicability for real life applications.
\section*{RESULT AND DISCUSSION}
The efficiencies of TEG for different temperature range, made up thermoelectric materials $Bi_2Te_3$, $Sb_{2}Te_{3}$, $PbTe$, $TAGS$, $CeFe_{4}Sb_{12}$, $SiGe$ and $TiO_{1.1}$ are depicted in FIG.2. 
\begin{figure}[ht!]
\centering
\includegraphics[width=3in]{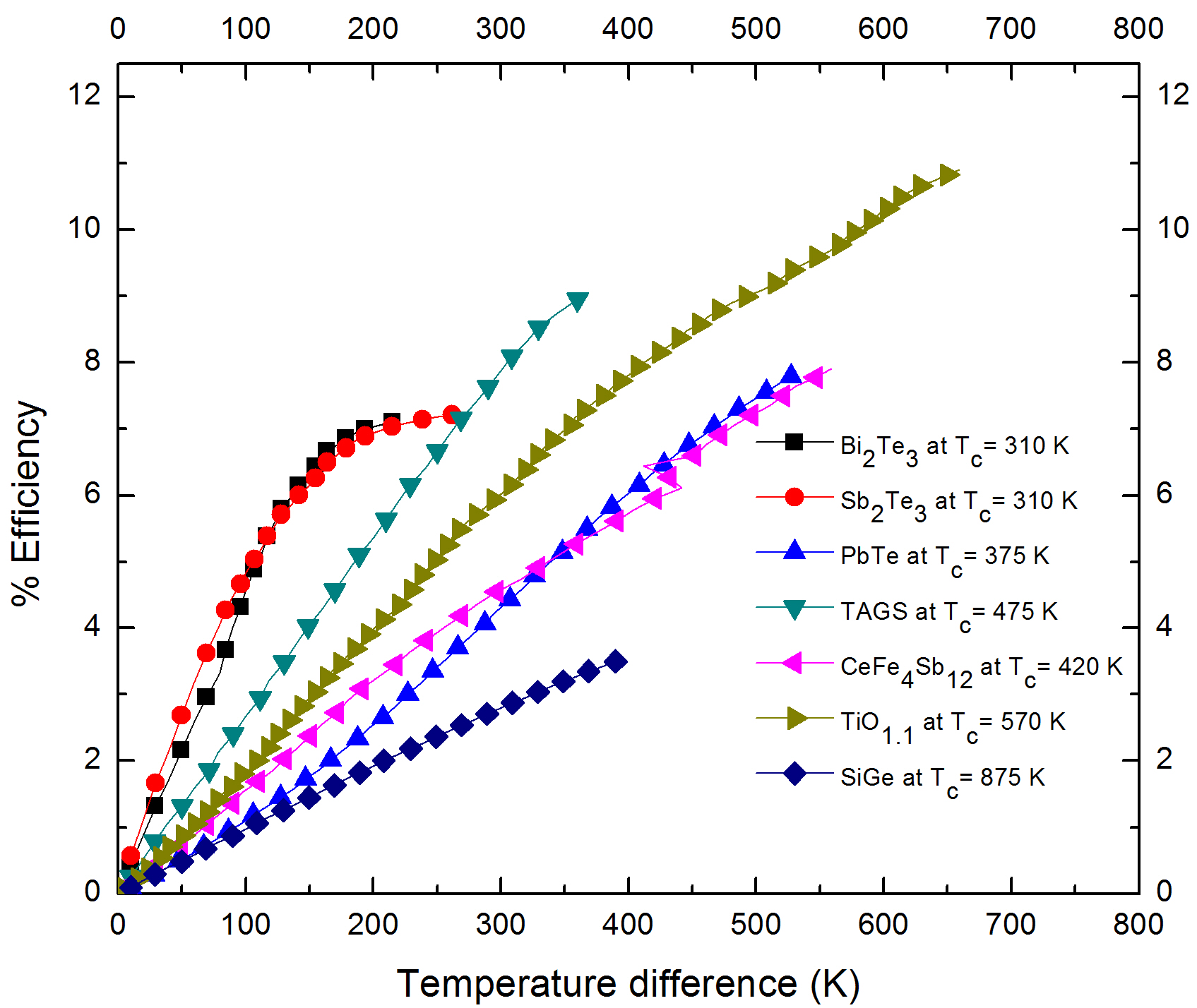}
\caption{Graph showing efficiency of TEG using different thermoelectric materials v/s temperature difference, keeping fixed cold side temperature}
\end{figure} The progressive increment in efficiency with a change in temperature difference is shown in FIG.2. Here, the temperature difference in abscissa denotes the arithmetic difference in temperature between two ends of the sample, keeping cold end temperature fixed with varying hot end temperature. Referring graph, the efficiency plot for $Bi_2Te_3$ and $Sb_{2}Te_{3}$ are nearly same, which tells that any one of them is worth to use, depending on the availability of resources. Another noticeable point is that the efficiency of TEG made up of $Bi_2Te_3$ and $Sb_{2}Te_{3}$ is getting almost saturated after 550 K. So, this also provides us information regarding the upper limit of working temperature. For hot end temperature up to 900 K, we can see that $PbTe$ and $CeFe_{4}Sb_{12}$ are having comparable efficiency $\sim$ 7\% to 8\%. This can be validated by the result reported by the Z. H. Dughaish\cite{Dughaish}. The efficiency of TEG using $TAGS$ as thermoelement is $\sim$ 9\%, which is quite more, however because of low sublimation temperature of $TAGS$ restrict their use. Now, from FIG.2, an inference can be drawn that efficiency of TEG using single material is limited up to 9\%. So, now we look for the possibility of enhancing the efficiency by using hybridization technique.\par The calculation is done for a hybrid module composed of $Bi_{2}Te_{3}$ and $PbTe$ for a temperature range of 310 K-500 K and 500 K-900 K, respectively, where selection of temperature range is done by considering compatibility factor. Using article (15), we have chosen the best suitable interface temperature, where compatibility factor of both materials is almost matching at temperature 500 K. This working temperature will yield close to maximum efficiency for TEG. The organization of calculation is as follows: first- $Bi_{2}Te_{3}$ has been considered for a temperature range of 310 K-500 K, second- $PbTe$ has been considered for 500 K-900 K. From FIG.2, the efficiency of TEG made up of single thermoelectric material is used to calculate the overall efficiency of hybrid TEG module. Considering firstly, TEG made up of $Bi_{2}Te_{3}$ for temperature range of 310 K- 500 K, the calculated efficiency is $\eta_{overall_1}= 6.56\%$. Similarly, one can calculate the efficiency of TEG made up of $PbTe$ in the temperature range of 500 K - 900 K. For doing this one can use Eq.(5) as a mathematical tool,
\begin{equation}
\eta_{2}=\frac{\eta-\eta_{1}}{1-\eta_{1}}
\end{equation} where, $\eta_{2}$ denotes the efficiency of TEG for a temperature range of 500 K to 900 K. $\eta$ and $\eta_{1}$ are efficiencies for a temperature range of 375 K-900 K and 375 K-500 K, respectively. The calculated efficiency of TEG thus obtained for $PbTe$ as thermoelement is $\eta_{overall2}=6.18\%$. Using FIG.2 and Eq.(5), we can calculate the theoretical efficiency of TEG for any temperature range. Using $\eta_{overall1}= 6.56\%$ and $\eta_{overall2}=6.18\%$ in Eq.(3), we got $\eta_{overall}= 12.33\%$. Here, we are presenting the individual and overall efficiency of $Bi_{2}Te_{3}$ and $PbTe$ in Table 1.
\begin{table}
\centering
\caption{ Efficiency obtained for different constituent of segmented thermoelectric generator and overall efficiency of entire module}
\begin{tabular}{|p{2cm}||p{1cm}|p{1cm}|p{2cm}|p{3cm}|}
\hline
$\text{Materials}$&\text{T$_{h}$ K}&\text{T$_{c}$ K}&\text{Calculated\%} \\ \hline
$Bi_{2}Te_{3}$ &585 &310 &7.10 \\ \hline
$PbTe$ &900 &375 &7.8 \\ \hline
$Hybridized$ &900 &310 &12.3 \\ \hline
\end{tabular}
\end{table}
In tabulation form, we can see that individual efficiency of each stacked layer is around 7\% to 8\% but after hybridizing both layers, efficiency shoots up to 12.3\%. We have plotted above-obtained results in graphical form in FIG.3.
\begin{figure}[ht!]
\centering
\includegraphics[width=3in]{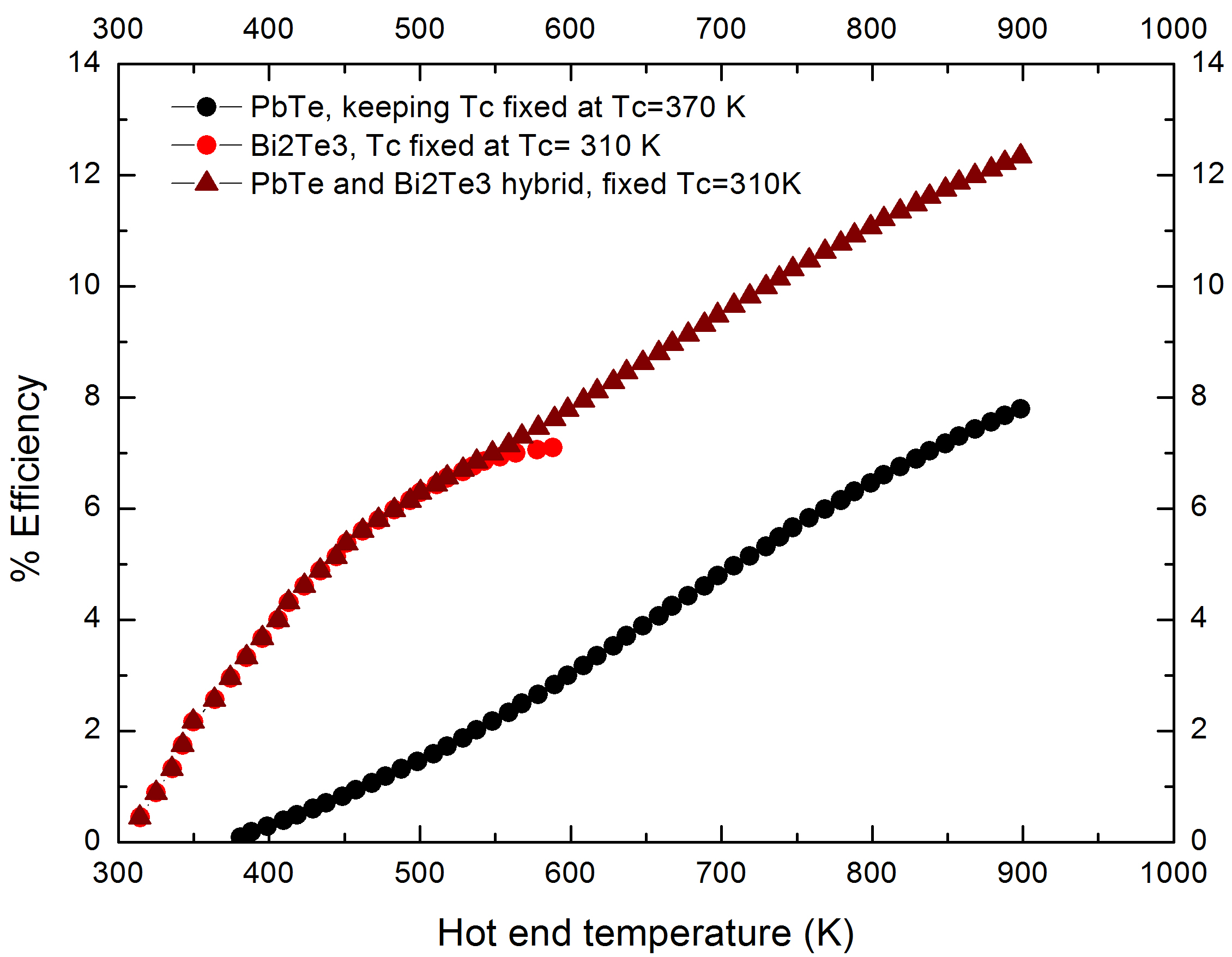}
\caption{ Efficiency v/s increasing hot end temperature, keeping lower temperature fixed}
\end{figure}
Here we are presenting variation in efficiency of the hybrid module with varying hot end temperature (keeping cold side temperature fixed). So, this graphical presentation will help us for calculating efficiency of any TEG setup installed between temperature range of fixed $T_{c}=310 K$ and variable hot end temperature.\par Proceeding with the same procedure as explained above for $Bi_{2}Te_{3}$ and $PbTe$, we are here with different combinations of above-mentioned materials. The aim is to produce different alternatives, in order to fulfill our need during the scarcity of one resource. So, here we are presenting cumulative efficiency of segmented TEG in Table 2 as well as in FIG.4.
\begin{table}
\centering
\caption{Cumulative efficiency obtained by different constituent of segmented thermoelectric generator in their operating temperature range}
\begin{tabular}{|p{1.5cm}|p{1.5cm}|p{1.5cm}|p{1.5cm}|p{1.2cm}|}
\hline
$\text{Material 1}$&\text{Range}&\text{Material 2}&\text{Range}&\text{Overall\%} \\ \hline
$Bi_{2}Te_{3}$ &310-425 K &$CeFe_{4}Sb_{12}$ &425-975 K &11.88\% \\ \hline
$Bi_{2}Te_{3}$ &310-510 K &$PbTe$ &510-900 K &12.33\% \\ \hline
$Bi_{2}Te_{3}$ &310-470 K &$TAGS$ &470-830 K &14.22\%\\ \hline
$Sb_{2}Te_{3}$ &310-500 K &$CeFe_{4}Sb_{12}$ &500-975 K &13.30\% \\ \hline
$Sb_{2}Te_{3}$ &310-510 K &$PbTe$ &510-900 K &12.90\% \\ \hline
$Sb_{2}Te_{3}$ &310-475 K &$TAGS$ &475-830 K &15.00\%\\ \hline
$Bi_{2}Te_{3}$ &310-570 K &$TiO_{1.1}$ &570-900 K &17.20\% \\ \hline
\end{tabular}
\end{table}
In Table 2, temperature range is decided by considering compatibility factor. The temperature at which compatibility of both materials are matching is taken as the best suitable interface temperature. Using the above technique of stacking two layers in series connection, we can see a significant increment in overall efficiency. So, if ignoring $TAGS$ as thermoelement in high temperature range (because of low sublimation temperature), the module of $Bi_{2}Te_{3}$ and $PbTe$ seems to be a viable option. The above mentioned combinations are plotted in graphical representation between the cumulative efficiency of stacked TEG and the variable hot end temperature (keeping cold end temperature fixed) as shown in FIG.4.
\begin{figure}[ht!]
\centering
\includegraphics[width=3in]{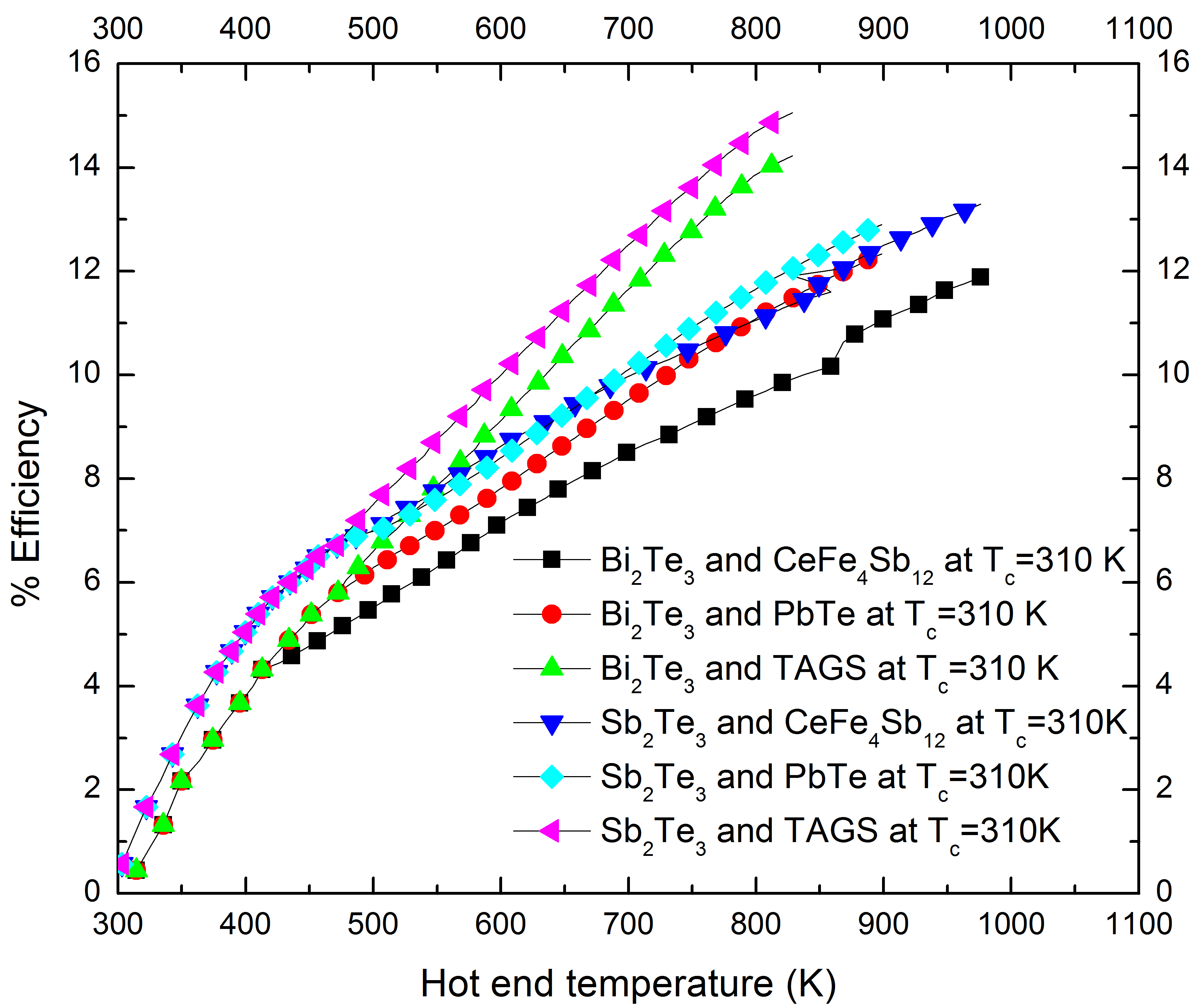}
\caption{ Cumulative efficiency v/s increasing hot end temperature for different materials, keeping lower temperature fixed}
\end{figure}
Table 2 and FIG.4 are showing the different possibilities of combining different available materials for the hybrid TEG. From FIG.4, we can conclude that after stacking two layers of thermoelectric material the efficiency goes up as well as it broadens our working temperature range. So, in the case of an automobile, where hot end temperature goes up to 900 K, this tabulated combination of different materials can be treated as a better alternative. Here, we have selected such type of materials whose melting point is higher than 1000 K. The purpose behind this selection is to know the efficiency of such TEG, which can be installed in an automobile. We already explained about efficiency and suitability of hybrid module of $Bi_{2}Te_{3}$ and $PbTe$ in an automobile. Rejection of other materials is based on their individual limitations such as $TAGS$ is having low sublimation temperature and $CeFe_{4}Sb_{12}$ is having very low strength (brittle in nature). So, these limitations restrict their use in an automobile because of heavy vibrational environment.\par From FIG.4, we can clearly notice that the efficiency, as well as working temperature of TEG, got increased. Till now, we have discussed efficiency calculation for automobile application where working temperature range is limited to 1000 K. If our aim is to use this TEG in a higher temperature range such as in spacecraft\cite{Spacecraft} and steel industry\cite{Okinaka}, there is a need to investigate materials having a good figure of merit as well as can sustain in high-temperature limit. For that, we found out $TiO_{1.1}$ and $SiGe$ as two different materials which can sustain at more than 1000 K. So, $TiO_{1.1}$ and $SiGe$ can be used in hybrid TEG for the upper-temperature region. We are calculating the efficiency of TEG made up of hybridization of $Bi_{2}Te_{3}$, $PbTe$ and $SiGe$ as thermoelements. Here, $Bi_{2}Te_{3}$, $PbTe$ and $SiGe$ are being stacked in low, medium, and high-temperature range, respectively. The interface temperature between $Bi_{2}Te_{3}$ and $PbTe$ is 510 K and between $PbTe$ and $SiGe$ is 890 K, which is decided by the matching value of $z\bar{T}$ for both materials. The addition of $SiGe$ as an extra thermoelement in previously explained $Bi_{2}Te_{3}$ and $PbTe$ module leads to raising the efficiency from 12.33\% to 15.24\%. Here we are facing a serious problem with compatibility of $SiGe$ with $PbTe$ because compatibility factor of $SiGe$ ($u$ $\sim$ 1) is significantly less than $PbTe$($u$ $\sim$ 2)\cite{Snyder2004}. This problem may reduce efficiency. The second problem is concerned with increasing number of stacked layers because it leads to increase in energy losses due to more numbers of interface contact surfaces. This decrease in efficiency will lead to the uneconomical design of TEG. Now, we have another alternative as $TiO_{1.1}$. So, we can make hybridization of $Bi_{2}Te_{3}$ and $TiO_{1.1}$. As we know that, for hybridization of layers, we need to know the interface temperature and that temperature will be deduced by considering compatibility factor. Using temperature-dependent thermoelectric material's properties\cite{Okinaka} of $TiO_{1.1}$, we have calculated compatibility factor with varying temperature, which is shown in FIG.5. We can see that at temperature 570 K, the compatibility factor of $Bi_{2}Te_{3}$ is $\sim$1 and for $TiO_{1.1}$ is $\sim$0.83. Since, we have similar values for compatibility factor of $Bi_{2}Te_{3}$ and $TiO_{1.1}$. So, now calculation of efficiency is done for a temperature range of 290 K to 1225 K using $Bi_{2}Te_{3}$ and $TiO_{1.1}$ as a hybrid module. This gives us an enormous efficiency of 17.20\%. Although, we have achieved a significant efficiency but it can be increased by taking the interface temperature at matching compatibility factor value. From FIG.5, we can see that the graph for compatibility factor of both materials are going to intersect on extrapolating the temperature to 580 K. This gives a slight increment in efficiency to 17.25\% at matching compatibility factor of $\sim$0.9. So, this efficiency can be treated as the best efficiency. Since the number of stacked layers are reduced to two, consequently it will reduce interface contact losses. Hence, we can achieve experimentally good efficiency. Another advantage of oxide is its capacity to prevent from rusting by reacting with atmospheric air.

\begin{figure}[ht!]
\centering
\includegraphics[width=3in]{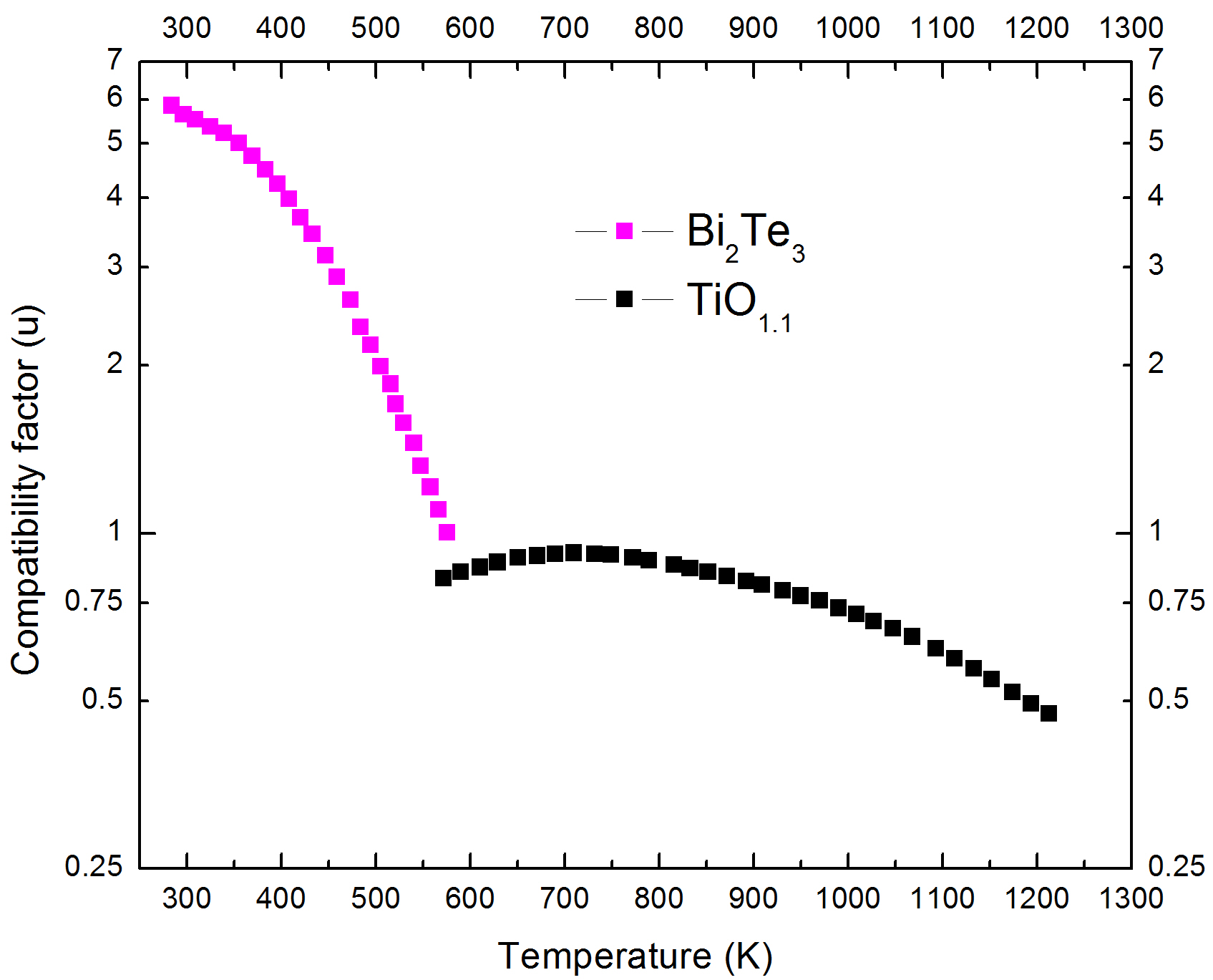}
\caption{showing compatibility of $Bi_{2}Te_{3}$ and $TiO_{1.1}$ with temperature of one end of sample}
\end{figure}

If our aim is to find out the efficiency of TEG made up of $Bi_{2}Te_{3}$ and $TiO_{1.1}$ for different hot end temperature (keeping cold side temperature fixed at $T_{c}=310 K$) then we can prefer FIG.6. In this figure, we are presenting the cumulative efficiency of TEG with varying hot end temperature. This type of curve can be used to know the efficiency for low to high-temperature range application site. For example, in automobile case for $T_{h}=900 K$ efficiency will be $\sim 13\%$ and for spacecraft as well as steel industry case for $T_{h}=1200 K$ efficiency will be $\sim 17\%$ (keeping fixed $T_{c}= 310 K$). This provides us diverse temperature range to find out the efficiency of TEG installed in different application site.
\begin{figure}[ht!]
\centering
\includegraphics[width=3in]{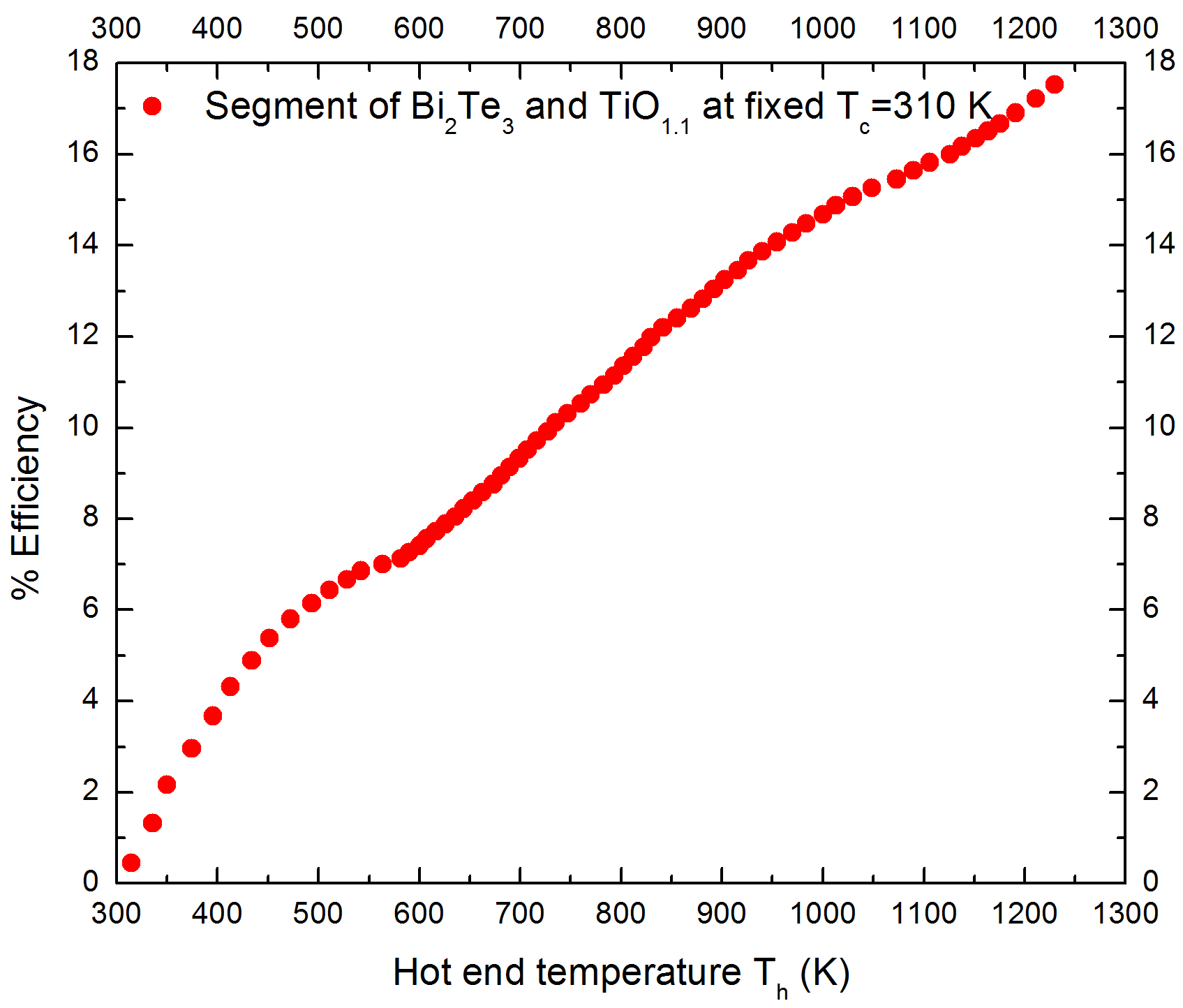}
\caption{showing efficiency by segmentation of $Bi_{2}Te_{3}$ and $TiO_{1.1}$ with hot end temperature, keeping cold side temperature fixed}
\end{figure} Now, finally we came with different materials and respective efficiency for different applications which include automobile, spacecraft, and steel industry.\par It is clear from the above discussion that, by using the temperature dependent data of $z\bar{T}$, one can calculate the efficiency of a TEG operating in any temperature range. Finally, it will help in selecting a proper thermoelectric material for a particular application. Considering automobile case, where the upper temperature is governed by the type of engine installed in that vehicle (keeping other parameters such as engine power, velocity, road condition, etc., constant). So, for higher exhaust port temperature we can install hybrid TEG module which can sustain at a higher temperature as well as have better efficiency. In India, steel industry consumes around 8200 MW power\cite{Firoz}, however, the recovery ratio of outflow heat is only 17\% in enthalpy basis and 25\% in exergy basis. There is still large room for exhaust heat recovery in the steel industry. Suppose we are going to install a TEG in steel industry, where exhaust gases leaving the furnace have minimum temperature is around 1200 K. Since, hot end temperature is governed by exhaust flue gases but cold end temperature depends on many parameters such as length of sample, environmental conditions, energy losses, heat sink conditions, heat-exchanger installed, etc. So, it's a tough and challenging task to achieve a fixed cold end temperature. Thus in FIG.7, we have estimated the temperature dependent efficiency of TEG with keeping fixed hot end and variable cold end temperature. Here, absolute temperature of the cold end can be calculated by subtracting the temperature difference from the temperature of hot end.

\begin{figure}[ht!]
\centering
\includegraphics[width=3in]{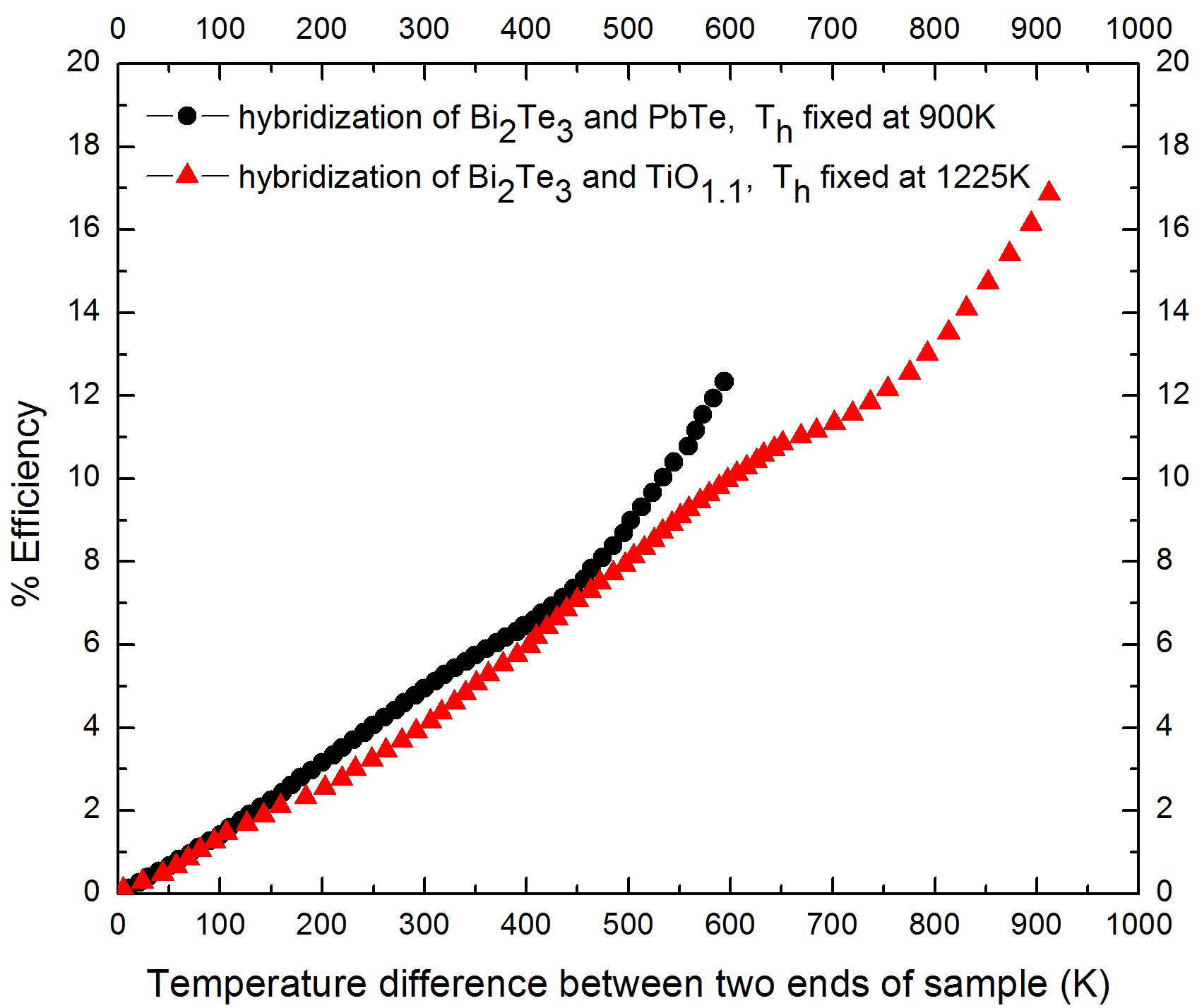}
\caption{showing efficiency combination of $Bi_{2}Te_{3}$ and $PbTe$ segment and $Bi_{2}Te_{3}$ and $TiO_{1.1}$ v/s temperature difference across sample, keeping higher temperature fixed}
\end{figure}

FIG.7 shows two curves, one is for combination of $Bi_{2}Te_{3}$ and $PbTe$ and other is for combination of $Bi_{2}Te_{3}$ and $TiO_{1.1}$ between a fixed hot end temperature and variable cold end temperature. So, as per requirement, we can opt for different curves. Suppose in case of automobile $T_{h} \sim 900 K$ and considering a typical value of $T_{c}\sim 600 K$. The efficiency for $\Delta T= 300 K$ is comes out to be $\sim$ 4\% and if we consider $T_{c} \sim 400 K$ then the efficiency for $\Delta T= 500 K$ is comes out to be $\sim$ 8\%. Similarly, for steel industry keeping $T_{h}$ = 1225 K and considering a typical cold end temperature $T_{c}$ = 600 K. So, for $\Delta T$ = 625 K, we can calculate the efficiency of TEG using FIG.7 as 10\%. This shows the versatility of FIG.7 that we can use it for any temperature range. Since our result obtained for the efficiency of TEG module made up of single material shows good agreement with Salzgeber et.al\cite{K Salzgeber}. Hence, the formulation for the hybrid module is equally valid. Such graph is expected to be very useful in deciding the efficient operating temperature range for TEG. These types of curves play an equally important role for design engineers as it directly provides information about efficiency by knowing temperature difference between hot and cold end. It also helps to judge the best possible temperature range for maximizing efficiency of a TEG due to various influencing operating conditions.

\section*{CONCLUSION}

We studied the applicability of $Bi_2Te_3$, $Sb_{2}Te_{3}$, $PbTe$, $TAGS$, $CeFe_{4}Sb_{12}$, $SiGe$ and $TiO_{1.1}$ as thermoelectric materials in fabrication of TEG by calculating their efficiency in temperature range of 310 K to 1200 K. We found out approximately $\sim$7\% efficiency of TEG made up of individual materials in their respective temperature range. Enhancement of efficiency from $\sim$7\% to $\sim$15\% is observed by hybridizing different combination of above mentioned materials for temperature range of 310 K to 900 K. We calculated the compatibility factor to judge the interface temperature of $Bi_{2}Te_{3}$ and $TiO_{1.1}$ and compatibility factor value found out $\sim$1 and $\sim$0.83, respectively at 570 K. Based on different properties like compatibility factor, sublimation temperature, and strength or brittleness, we found that the combination of $Bi_{2}Te_{3}$ and $PbTe$ can be a better option for automobile application. The combination of $Bi_{2}Te_{3}$ and $PbTe$ gives an efficiency of $\sim$12.33\% in temperature range of 310 k to 900 K. Similarly, in higher temperature range application such as spacecraft and steel industry, combination of $Bi_{2}Te_{3}$ and $TiO_{1.1}$ is considered for temperature range of 310 K to 1200 K, which gives an efficiency of $\sim 17\%$. Furthermore, we have shown the progressive effect in efficiency with varying cold end temperature (keeping fixed hot end temperature) of TEG module.

\bibliographystyle{unsrt}  


\end{document}